\documentclass[a4paperfleqn,usenatbib]{mnras}

\usepackage[T1]{fontenc}
\usepackage{ae,aecompl}
\usepackage{float}
\usepackage{graphicx}   
\usepackage{color}
\usepackage{amssymb}
\usepackage{amsmath}
\usepackage{enumitem}
\usepackage{booktabs} 

\title[Blazhko stars in M3]{ Blazhko type fundamental-mode RR Lyrae stars in the globular cluster, M3}
\author[J. Jurcsik et al.]{{J. Jurcsik$^{1}$\thanks{E-mail:jurcsik@konkoly.hu}}\\
$^{1}$Konkoly Observatory, H-1121 Budapest, Konkoly Thege Mikl\'os \'ut 15-17., Hungary\\
\\
}
\begin{document}
\date{Accepted 2019 September 4. Received 2019 August 26; in original form 2019 July 19 }
\pagerange{\pageref{firstpage}--\pageref{lastpage}} \pubyear{2019}
\maketitle

\label{firstpage}

\begin{abstract}
Blazhko stars from the extended $BVI_{\mathrm C}$ time series of RRab stars in the globular cluster, M3, are analysed. This is the largest sample of Blazhko stars with full details of their Blazhko properties in a homogeneous stellar system. Blazhko periods and light-curve solutions are determined/estimated for 83 fundamental-mode variables. 
The lack of phase modulation in Oosterhoff-type~II stars and the diminishing relative strength of the modulation in long-period Oosterhoff-type~I stars are the regular tendencies found between the pulsation and modulation properties of Blazko stars. Similarly to previous results, no modulation of the longest-period stars is detected. The onset of the modulation in a previously regular RRab star, and the similar distribution of modulated and non-modulated stars imply that the modulation is a temporal property of RRL stars, which may occur at any time in any RRab star except the coolest ones. Comparing the modulation periods in M3 and in other samples of Blazhko stars, the  mean log$(P_{\mathrm {mod}})$ value is found to depend on  the metallicity of the system. The separation of the temperature- and radius-change induced variations supports our previous finding that the photometric radius variation does not show any modulation. The pulsation-averaged mean brightness and temperature of Blazhko stars are found to be larger in the large-amplitude phase of the modulation than in the small-amplitude phase. The larger the amplitude of the modulation, the larger changes of the mean parameters are detected.
\end{abstract}

\section{Introduction}

The RR Lyrae (RRL) variables, as  horizontal-branch stars with very similar internal structures, represent a unique check-point in the evolution of low mass stars after the red giant branch phase. As their H-burning time on the main sequence is very long, they are relics of the original old population of star formation in the galaxies (for reviews see \citealt{ca} and  \citealt{csbook}). 

However, RRL stars still exhibit some incompletely understood phenomena: the Blazhko effect  (the amplitude and/or phase variations of the pulsation; \citealt{blazhko} and \citealt{shapley}), the anomalous period ratios of some double-mode stars, the mysterious additional frequencies of the variables (e.g., \citealt{mo15,smo15,netzela,netzelb,m3mod,overtone}), as well as  the  period-doubling phenomenon \citep{szabo}. Taking into account that RRL stars are key objects for the studies of the horizontal branch, the distance of remote systems, and chemical composition analysis revealing different generations of old stellar populations, hence, any property of RRL stars that we cannot explain hold uncertainties on the results of any other study based on these stars.

Due to the efforts of space missions \citep[$CoRoT$, $Kepler$:][]{corot,kepler} and ground based projects \cite[Konkoly Blazhko Survey, TAROT, OGLE:][]{stat,tarot,ogleIV} many new aspects were uncovered about Blazhko-modulated stars in recent years, however the origin of the phenomenon has remained undisclosed. Taking into account the fact that the modulation appears in about 40 percent of the RR Lyrae stars in the Galactic bulge \citep{prudil}, 50 percent in the field \citep{stat,be}, and an even larger fraction can be affected in some globular clusters \citep{m53,ngc2808}, the lack of understanding of the phenomenon cannot be neglected.

Using the OGLE data of more than 8000 stars, the comparison of Blazhko and mono-periodic RRL star of the Galactic Bulge has led to the detection of some systematic differences between their period distributions and light-curve characteristics, e.g., amplitude, rise time, Fourier amplitude ratios and phase differences \citep{prudil}. The metallicity distribution of the Galactic bulge RRL stars covering  larger than 2 dex full range is sharply peaked at [Fe/H]$=-1.0$ \citep{piet12,prudil,dekany}. Most probably it contains  multiple old stellar populations \citep{piet15,dekany}. As the shape of the light-curve of an RRL star is determined  by its physical parameters completely,  the interpretation of the differences between Blazhko and non-Blazhko RRL stars is not unambiguous in an inhomogeneous sample. 

A reliable comparison of the physical properties of Blazhko and mono-periodic RRL stars is only possible using Globular cluster (GC) data. GCs are frequent and popular targets of RRL studies, because they are rich in this  type of variables and they provide relatively homogeneous samples compared to the other targets of massive photometric campaigns (e.g., Galactic field, Galactic bulge, satellite galaxies). The observations of these compact groups of stars, where the relative values of the physical parameters can be determined with uniquely high precision, provide an outstanding opportunity for an in-depth study of their stellar populations.

The extended  $BVI_{\mathrm C}$ photometry and the non-calibrated time series of the most crowded stars published in \cite{m3data} enable us to analyse the complete Blazhko RRL sample of the M3 globular cluster. This is the largest multi-colour photometric data set of a homogeneous group of Blazhko stars ever observed, which is suitable to detect any connection, if it exists, between the properties of the modulation (the variations of the pulsation changes during the modulation cycle)  and the physical parameters of the variables. In this paper fundamental-mode RRL stars are studied, the Blazhko properties of the  overtone and double-mode variables in M3 were discussed in \cite{m3mod,overtone}. Some preliminary results were already shown in \cite{rrconf}.

\section{Data and light-curve analysis}

Time series data were obtained for the complete sample of RRab stars in M3  in 2012. $B, V, I_{\mathrm{C}}$ magnitudes for about the two third of the variables and relative $b$ fluxes for the most crowded variables, which magnitude calibration failed, were published in \cite{m3data}, together with the full details of the photometric process. 

Using these data, RRab stars showing Blazhko modulation are identified and their modulation periods are determined. The photometric time-series data are fitted with Fourier sum of the pulsation ($k\times f_0$) and the modulation components  ($k\times f_0 \pm n\times f_{\mathrm m}, k \geq 0; n \geq 1$). The modulation of some of the Blazhko stars are multiple periodic. The linear-combination terms of the different modulation components also occur in these cases. 

The relevant pulsation and modulation components are determined star by star in the course of a successive pre-whitening process of the $B$ and $V$ light curves using the program package {\textsc{mufran}} \citep{mufran}.  As the signal-to-noise properties are worse in the $I_{\mathrm{C}}$ band than in the $B$ and $V$ bands, the  $I_{\mathrm{C}}$ time series data were not used in the course of the frequency identification. After determining the frequency components appearing in the $B$ and $V$ data sets, a non-linear fitting algorithm \citep{nl} is applied separately on the $B$ and $V$ data  to refine the values of  $f_0$ and $f_{\mathrm m}$. Finally, the average values of the results obtained for the $B$ and $V$ data are accepted as the $f_0$ and the $f_{\mathrm m}$ frequencies of the star, and the full light-curve solution is determined as a linear fit to the data using the appropriate combination of these locked frequencies in each band ($B, V, I_{\mathrm{C}}$).  

Based on the difference between the results obtained for the two bands and on trials using different modulation components, the accuracy of the modulation  periods  is estimated  to be better than 1 d for periods shorter than 100 d and  about $1-10$ d for longer modulation  periods. However, the modulation periods  are less certain for long modulation cycles and for variables showing complex, and multiple-periodic modulations. The actual values of the modulation frequencies depend on the selection of the frequency components involved in the fit more strongly in these latter cases. The modulation periods of crowded variables  with non-calibrated photometric data are determined similarly, but their accuracy might be lower especially for noisy data and for small-amplitude modulations.

\begin{figure}
\centering
\includegraphics[width=8.9cm]{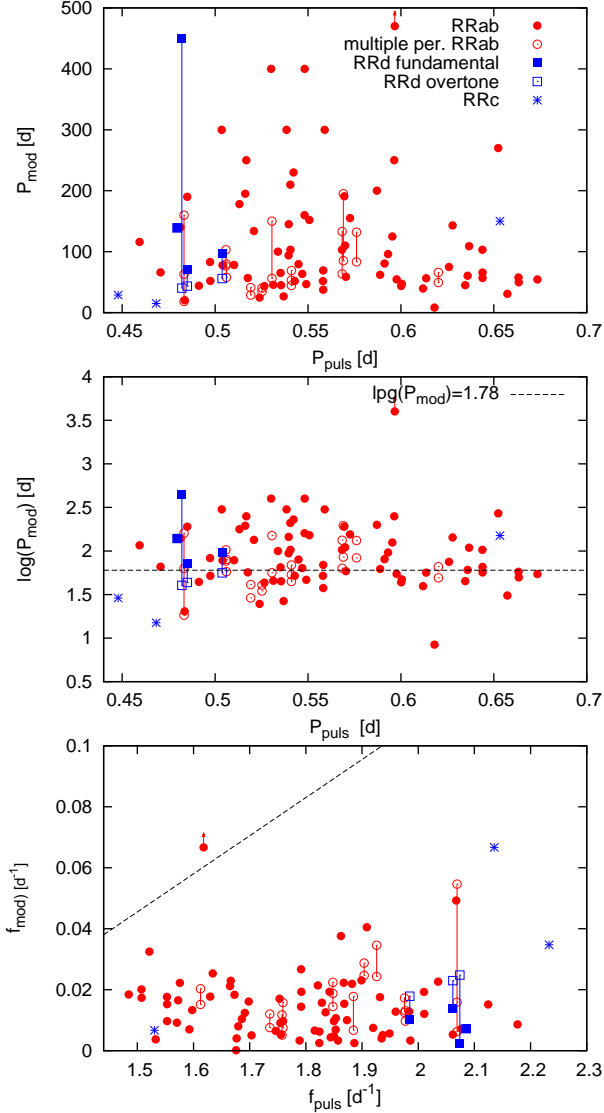}
\caption{Modulation  period versus pulsation  period (top panel), the same parameters but plotting the  $(P_{\mathrm {mod}})$ on a logarithmic scale (middle panel), and modulation frequency versus pulsation frequency (bottom panel) for Blazhko stars in M3 are shown. Data of overtone and double-mode stars are taken from \citep{overtone}. RRab stars with multiple modulations are denoted by open circles, and the modulations of the fundamental and the overtone modes of RRd stars are denoted by filled and open squares, respectively. The overtone-mode periods are fundamentalized. The different  modulation  periods/frequencies of RRd and multiple-modulated stars are connected by vertical lines. The position of the extreme-long modulation-cycle Blazhko star, V144, and the uncertain, extreme-large  modulation frequency of V134 are indicated by  vertical arrows in the top/middle and bottom panels, respectively.  The dashed line drawn in the middle and bottom  panels correspond to the mean log$(P_{\mathrm{mod}})$ value determined for the sample of 1628 Blazhko stars by \citet{marek}, and the possible upper limit for the modulation frequencies as found in \citet{j05}.   }
\label{per.fig} 
\end{figure}

The 2012 observations span over about a 200-day interval but data at the beginning and at the end of the observing season are sparse. Therefore, in order to refine modulation  periods longer than 100 d, if available, the observations obtained in 2009 \citep{oc} are added to the data set. However, because of strong  period and modulation-content changes and aliasing problems,  light-curve solution is obtained for the full $2009-2012$ data set only for some of the Blazhko stars.

Table~\ref{rrbl.dat} lists the Blazhko stars, their Oosterhoff (Oo) type, the pulsation and modulation periods determined from the 2012 data, and the refined periods using additional previous photometric observations for long Blazhko-cycle stars if a common light-curve solution is obtained  for the combined data.  OoII-type stars are typically brighter and of larger amplitude than OoI stars at the same period.  Here, the OoII status of Blazhko stars are determined based on the mean magnitudes \cite[published in][]{m3data} rather than on the position of the star in the period-amplitude Bailey diagram. The Oo~status of two Blazhko stars (V048 and V078) indicated by the amplitude and the mean brightness of the stars are controversial. This may originate from some photometric defect and/or from any bias on the amplitudes caused by the modulation. The Oo~status cannot be determined for variables with  non-calibrated light curves. 

\section{Modulation  periods}

The distribution of the modulation-period values does not show any convincing connection with the pulsation period values as documented in the top panel of Fig.~\ref{per.fig}. For completeness, the modulation properties of the M3 RRd and RRc stars analysed in \cite{overtone} are also displayed in the figure using fundamentalized pulsation period for the overtone mode.

Collecting a large sample of Blazhko stars (1628 stars)  \cite{marek} found that the distribution of the modulation periods is  log-normal  with log$(P_{\mathrm{mod}})=1.78(3)\pm 0.30$ mean  value. The $1\sigma$ uncertainty of the mean is given in parenthesis. This mean value is shown by dashed line in the middle panel of Fig.~\ref{per.fig}. The mean value of the Blazhko periods in M3 is larger than obtained for the inhomogeneous sample of stars  by \cite{marek} as can be seen in  Fig.~\ref{per.fig}. In M3, the mean log$(P_{\mathrm{mod}})$ is $1.91(4)\pm 0.39$ and $1.92(4)\pm 0.39$ for the total sample (RRab/RRd/RRc)  and for the fundamental-mode variables, respectively. The difference between the logarithmic mean values of the Blazhko periods in M3 and in the sample collected by \cite{marek} is around the $3\sigma$ uncertainty of the mean, indicating that the detected difference may be real.

\begin{figure*}
\centering
\includegraphics[width=18.cm]{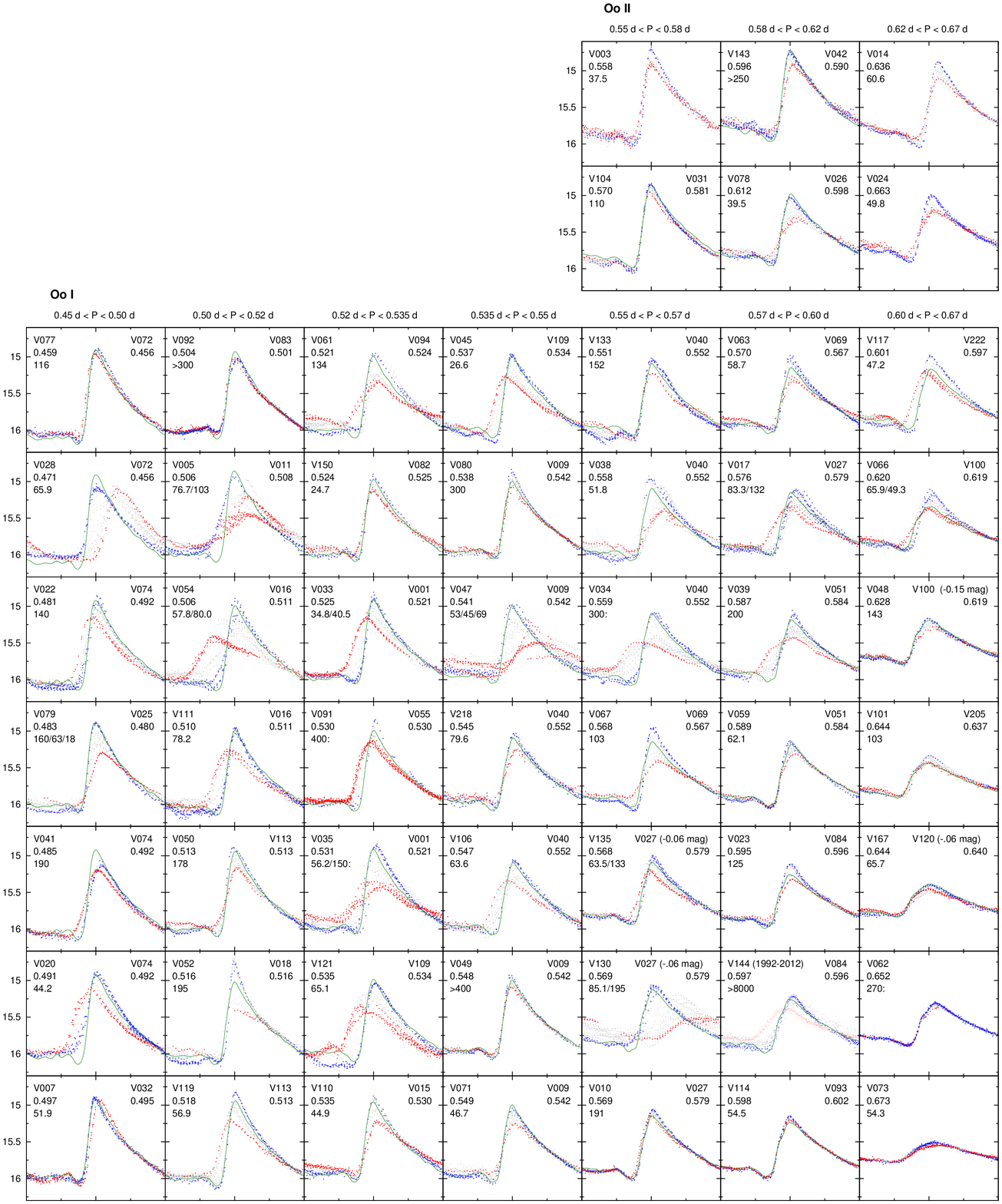}
\caption{Phased $V$ light curves of the magnitude-calibrated Blazhko RRab stars in M3. OoI- and OoII-type stars are plotted separately. The pulsation  period of the stars are increasing from the left to the right in the figure as indicated in the top of the columns. The smallest- and largest-amplitude phases of the modulation for each star are set in red and blue colours, respectively. The ID and the pulsation and the modulation  periods of the stars are given in the top left-side corner of the panels. For comparison, the green line shows the light curve of a similar  period/magnitude/colour stable RRab star, if such a star exists. The ID and the  period of these stable stars are given in the top right-side corner of the panels. }
\label{lc} 
\end{figure*}

\begin{figure*}
\centering
\includegraphics[width=18.cm]{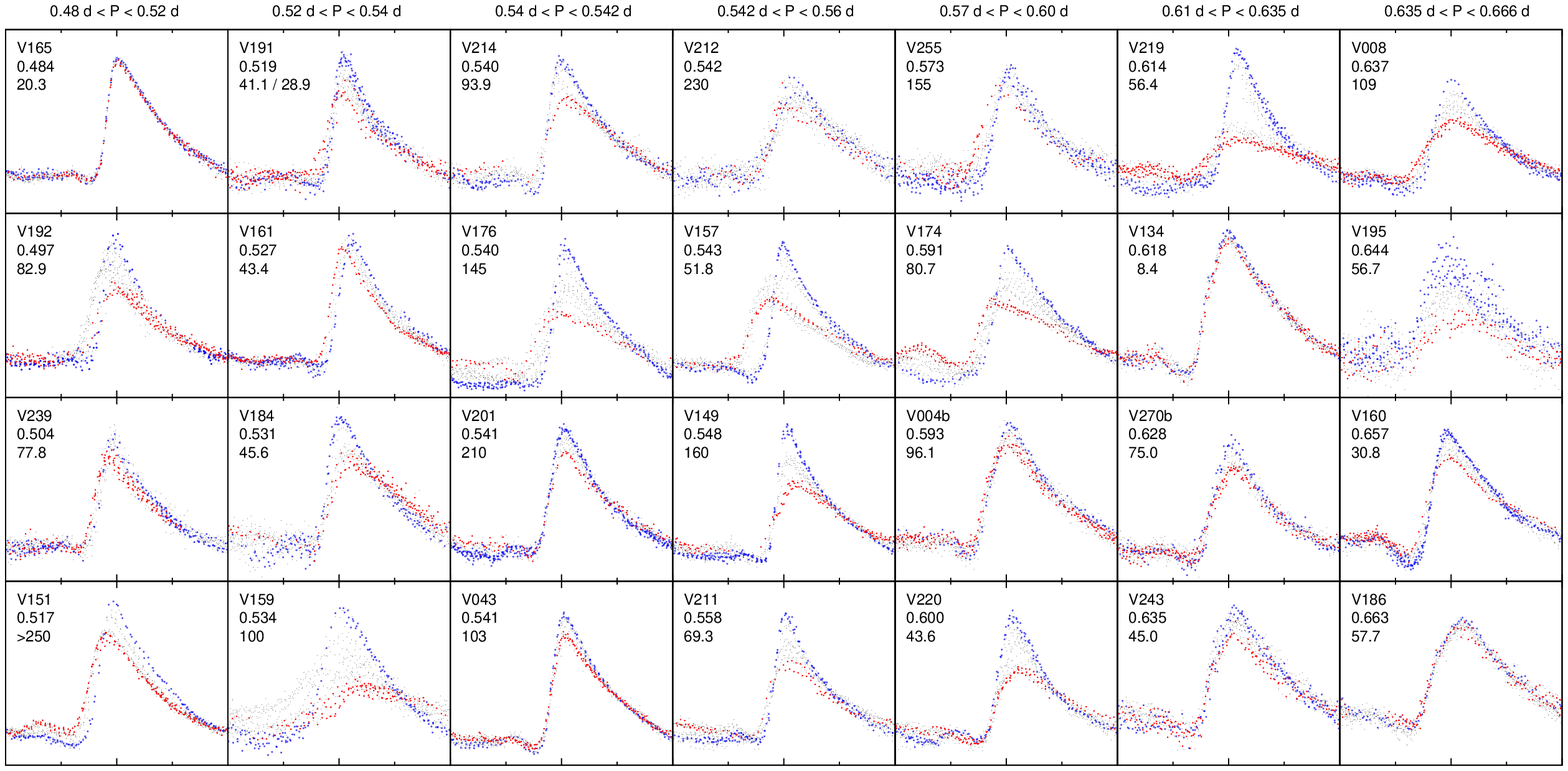}
\caption{Instrumental $b$ light curves of Blazhko stars with no standard magnitude calibration. The smallest- and largest-amplitude phases of the modulation are set in blue and red colours, respectively.} 
\label{lcb} 
\end{figure*}

A large fraction of the Blazhko stars used by \cite{marek} belongs to the Large-Magellanic Cloud (LMC) \citep{alcock}. The mean [Fe/H] of the LMC RRL stars is $-1.5$ \citep{lmc}, i.e., the LMC RRL stars have, on the average, similar metallicity as in M3. Therefore, we have  checked  what  the mean value of the log$(P_{\mathrm{mod}})$ is for the LMC sample alone. For the 731 LMC Blazhko stars given by \cite{alcock} it is $1.95(2)\pm 0.42$, which is identical with the result obtained for M3 within $1\sigma$. 

To reach 1.78 mean log$(P_{\mathrm{mod}})$ value for 1628 stars when nearly the half of the sample yields 1.95, requires that the mean of the other part of the sample has to be as small as 1.64.
 
Besides the LMC stars the sample used by \cite{marek} contained mostly Galactic field and bulge stars. The metallicity of the bulge peaks at [Fe/H=$-1.0$, and field stars are typically more metal-rich than variables in GCs and in the LMC.

  Based on these results, we conclude that the mean log$(P_{\mathrm{mod}})$ depends on the metallicity if large enough samples of Blazhko stars are considered. The more metal-poor a system, the larger the mean value of the Blazhko periods on a logarithmic scale.

Previously,  based on data of the Large- and Small-Magellanic Clouds, GCs and field stars it was found that the upper limit of the possible range of the modulation frequencies is increasing with increasing pulsation frequencies \citep{j05}. In order to check the validity of this result on the M3 data the modulation frequencies are plotted versus the pulsation frequencies in the bottom panel on  Fig.~\ref{per.fig}. The upper limit for the modulation frequencies detected in \cite{j05} is drawn in this plot for guidance. The M3 data are systematically below this upper limit and the slight tendency for an increase of the largest $f_m$ values with increasing $f_0$ is not convincing.

The modulation periods/frequencies do not depend on the brightness and colour of the variables.

\section{The light curves}\label{lc.sec}

\begin{figure*}
\centering
\includegraphics[width=17.cm]{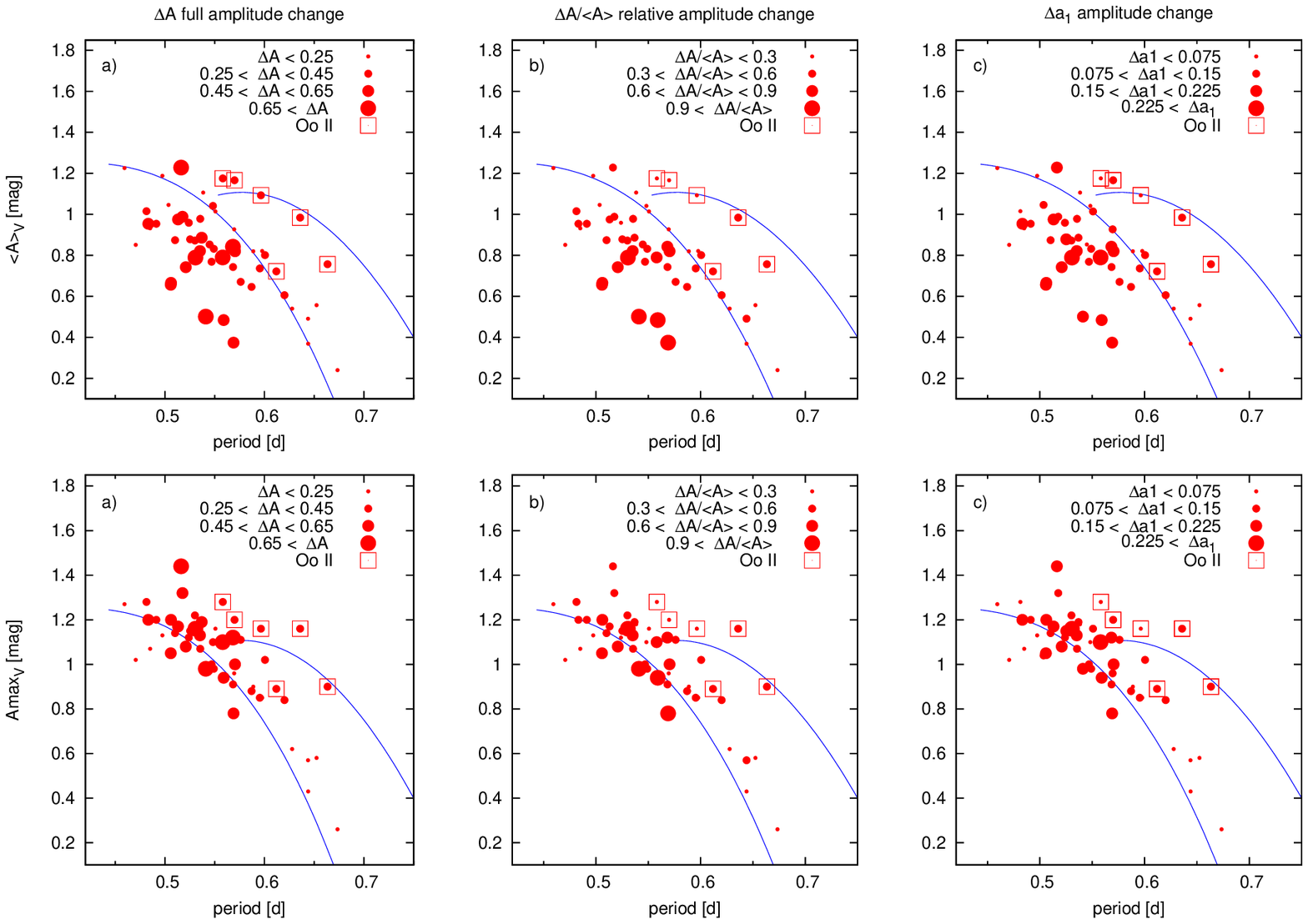}
\caption{Positions of Blazhko RRab stars in the   period-$V$ amplitude diagrams. The loci of OoI- and OoII-type stable light-curve RRab stars in M3 are indicated by blue lines. The amplitude corresponds to the mean and the maximum $V$ amplitudes of Blazhko stars in the top- and bottom-row panels, respectively. The positions of OoII Blazhko stars are marked by square symbols.
The strength of the amplitude modulation is indicated be the size of the symbol as given in the legends. Its determination is threefold: the full ranges of $a)$  the  changes of the peak to peak amplitude, $b)$ its relative value, and $c)$  the changes of the amplitude of the $f_0$ Fourier component during the Blazhko cycle are measured. }
\label{am} 
\end{figure*}

\begin{figure*}
\centering
\includegraphics[width=17.cm]{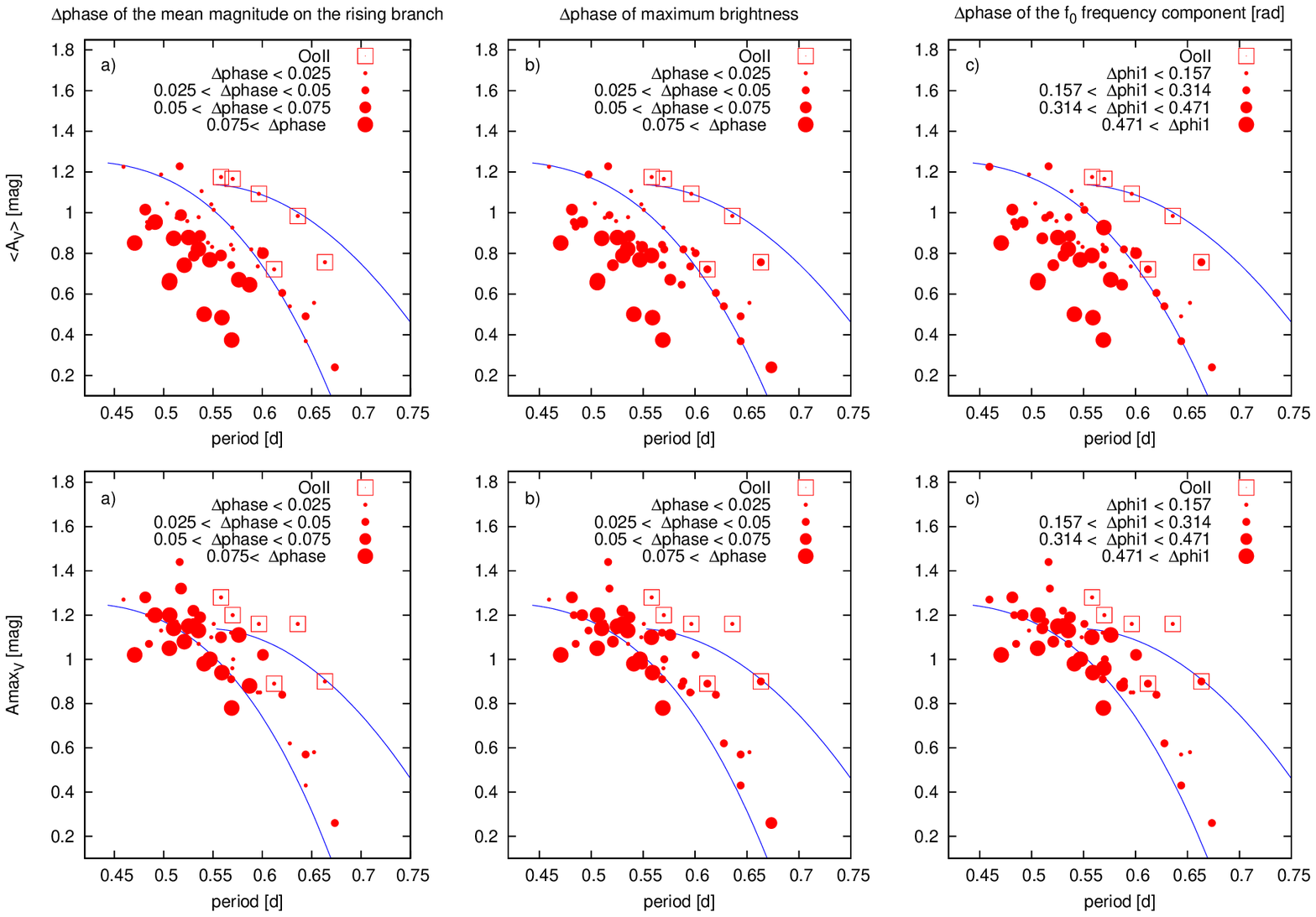}
\caption{The same  period vs mean-amplitude and period vs maximum-amplitude plots of Blazhko stars are shown as in Fig.~\ref{am}, but the symbols' size indicates the strength of the phase modulation here. The phase-modulation strength is determined threefold: the full ranges of  $a)$ the phase change of the position of the mean magnitude on the rising branch, $b)$ the phase change of the maximum brightness, $c)$ and the phase differences of the $f_0$ frequency  component in different phases of the modulation are used. }
\label{fm} 
\end{figure*}

The phased $V$ light curves for the Blazhko stars with magnitude calibrated time series data are shown in Fig.~\ref{lc}.  The light curves of the most crowded Blazhko stars with relative $b$ fluxes published in \cite{m3data} are transformed to an arbitrary magnitude scale matching the mean magnitude of the horizontal branch of M3, are shown in Fig.~\ref{lcb}.
In order to detect any tendency occurring in the distribution of the modulation properties easily, the pulsation  period is increasing from the left to the right in the figures (the period ranges are indicated in the top of the columns),  and the OoI- and OoII-type stars are plotted separately in Fig.~\ref{lc}.

For comparison, the light curve of a stable RRab star close in  period, brightness and colour to a given Blazhko star, if such a star exists in the M3 sample, is also displayed in Fig.~\ref{lc} using the data published in \cite{m3data}. The IDs and the periods of the Blazhko stars and their stable pairs are given in the top left-side and the top right-side corners of the light curves, respectively.

Looking at these figures, no straightforward conclusion on the distribution of the modulation properties can be drawn.  The modulated light curves show highly diverse characteristics. There are stars showing small- and large-amplitude modulations with different strength amplitude- and phase-modulation components in each part of the figures. The small-amplitude phase precedes the large-amplitude one in many phase-modulated light curves but there are examples for the opposite behaviour, too.
The modulation  properties of stars with very similar  period, brightness and colour, e.g., V022/V079 ($P_{\mathrm {puls}}=0.48$ d); V034/V038 ($P_{\mathrm {puls}}=0.56$ d); V039/V059 ($P_{\mathrm {puls}}=0.59$ d) can be quite different, and conversely,  similar-shape modulations of different pulsation- and modulation-period variables are observed, e.g. V045/V117 $P_{\mathrm {puls}}:0.54/0.60$ d, $P_{\mathrm {Bl}}: 27/47$ d and  V071/V023/V024 $P_{\mathrm {puls}}:0.55/0.59/0.66$ d, $P_{\mathrm {Bl}}: 47/125/50$ d.
 
One regularity that might be suspected is the lack of any strong modulation in the five longest-period ($P_{\mathrm {puls}}> 0.62$~d), smallest-amplitude  OoI-type stars.

Another possible systematic feature is the lack of phase modulation in the OoII sample.
The OoII-type Blazhko stars exhibit only small- or modest-amplitude amplitude modulation and negligible phase modulation. 
However, the small sample size (six stars) makes some doubts about the reality of this result. Other samples of well identified OoII Blazhko stars are needed to check the possible weakening/disappearance of the phase modulation in the OoII population.

The comparison  of the Blazhko stars with their non-Blazhko `twins' is not conclusive either. There are Blazhko stars with  similar-shape  light curves to their stable pairs in the largest-amplitude phase of the modulation, but the matching is not perfect. Small differences in the structure of the bump, the steepness of the rising branch and/or in the total amplitude are detected in these cases (e.g., V007, V033, V048, V071, V110, V119, V218). Moreover, there are also variables whose light curves in the large-amplitude phase of the modulation differ significantly from the shape of a normal light curve, see e.g., V005, V020, V028 V052, V117, V121, V130.

There are also stars with amplitudes smaller than normal in each phase of the modulation like V005, V028, V041. The modulation cycle of V041 is not fully covered by the 2012 observations, but checking the light curves of V041 in  all the previous CCD $V$ observations \citep{cc,bj,oc} this conclusion remains valid.

The shape of the light minimum is abnormally shallow and featureless in some stars (V005, V020, V022, V028, V034, V130, V159). All these stars exhibit strong phase modulation, but this is not always true inversely, the minimum phase is not featureless in each phase-modulated Blazhko star, see e.g. V117, V121, V174.

The light curves of Blazhko stars showing only weak modulation are close to normal in each phase of the modulation,  and vice versa, strongly deformed light curves are observed only if the modulation is strong.

\section{Amplitude and phase modulations}

The strength and the period-distribution of the amplitude- and phase-modulation components of the modulation  are investigated using different measures and showing them by the symbol sizes when plotting the positions of Blazhko stars in the period-amplitude Bailey diagram. The results are shown in Figs.~\ref{am} and ~\ref{fm} for the strength of the amplitude- and for the phase-modulation components, respectively. The construction of the Bailey diagram is twofold: either the mean- (top row in the figures) or the maximum-amplitude values (bottom row in the figures) of Blazhko stars are plotted. The ridge lines of the OoI and OoII populations defined by stable light-curve variables in M3 are drawn in the figures for comparison.

The strength of the amplitude modulation is quantified by the full ranges of $a)$  the peak to peak amplitude ($\Delta A$) change, $b)$ its relative value ($\Delta A/<A>$), and $c)$ the amplitude change of the $f_0$ Fourier component ($\Delta a_1$) of the light curve. 

Each plot in the top row of Fig.~\ref{am} shows that the mean amplitudes of Blazhko stars are significantly smaller than normal for a large fraction of Blazhko stars. The larger the amplitude of the amplitude modulation, the smaller the mean amplitude of the star, i.e., larger symbols are at systematically smaller amplitudes as normal.
 Looking at the distribution of the strength of the amplitude modulation on the Bailey diagrams constructed using the maximum amplitude (bottom-row panels in Fig.~\ref{am}) we see that variables with intense modulations shown by large-size symbols have either smaller or larger than normal maximum amplitudes, but there are some stars with similar maximum amplitude to the amplitude of non-modulated stars. Therefore, it can be  concluded that neither the mean nor the maximum amplitudes of Blazhko stars can be identified with the normal amplitude of the star automatically, and this is especially true for stars displaying large-amplitude modulation.


It can be also seen in each plot of Fig.~\ref{am} that the small-size symbols tend to be close to the position of the ridge lines of non-Blazhko stars. The exception here is V028, this star exhibits only a small amplitude change but its mean magnitude is significantly reduced due to its extreme large phase modulation.  

Fig.~\ref{am} also shows that the amplitude modulation of the longest-period OoI stars is small. This conclusion is also valid for the relative-amplitude change, therefore, this effect in not a bias caused by the overall amplitude decrease of the pulsation towards long periods.

\cite{prudil} showed that the difference between the mean amplitudes of Blahzko stars and of normal RRab stars decreases linearly towards longer periods. Taking into account that the amplitudes of Blazhko stars with small modulation amplitudes are more or less normal, and that the modulation amplitudes of longer-period Blazhko stars are small, the results obtained in M3 is in line with the trend detected in the bulge by \cite{prudil}.

To quantify the strength of the phase-modulation components we use three different measures: the full ranges of $a)$  the phase change of the position of the mean magnitude on the rising branch, $b)$ the phase change of the maximum brightness, $c)$ and the Fourier phase differences of the $f_0$ frequency component. The size of the dot symbols indicates the strength of the phase-modulation according to these measures in Fig.~\ref{fm}.

Each  diagram  in the top row of Fig.~\ref{fm} shows a tendency that towards larger mean amplitudes the strengths of the phase modulation is decreasing at a given period, and {\it vice versa} the larger the phase modulation, the smaller the mean amplitude of the variable, similarly to what was found for the amplitude modulation. 

 It was concluded from the inspection of the light curves in Sect.~\ref{lc.sec} that OoII-type stars do not display significant phase modulation.
The results shown in Fig.~\ref{fm} support this finding.

The positions of Blazhko stars in the Bailey diagram are more or less normal unless the amplitude of the phase modulation is large.  Concerning this conclusion, there is no significant difference between the results shown in the six panels of Fig.~\ref{fm}.

The Bailey diagrams using the maximum amplitudes (bottom row in Fig.~\ref{fm}) do not show any  additional special distribution for the strengths of the phase modulation.

\section{Multiple  periodic, complex modulations}

\begin{figure}
\centering
\includegraphics[width=8.5cm]{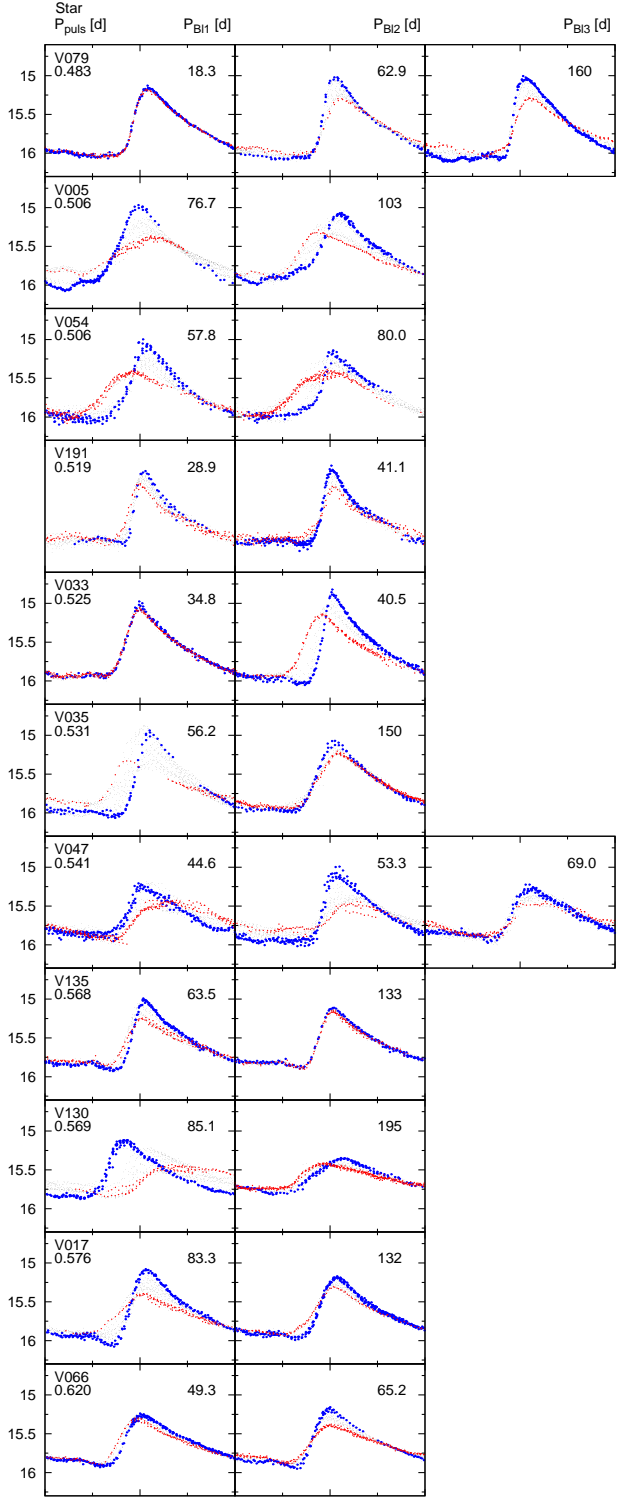}
\caption{The $V$ light curves (the $b$ flux curves for V191) of multiple-periodic Blazhko stars pre-whitened for one or the other modulation component(s) are shown. The 11 stars are plotted in the order of increasing pulsation  period from the top to the bottom, while their modulations are displayed in the order of increasing modulation  period from the left to the right for each star. The light curves in the largest and smallest amplitude phases of the modulations are set in blue and red colours, respectively.}
\label{bl2} 
\end{figure}

The modulation of 11 RRab stars cannot be described by using a single modulation frequency; two or tree modulation frequencies are needed for an appropriate fit to the data. 

After determining the complete Fourier solutions of the light curves for these stars, it becomes possible to separate the light-curve modulations connected to each modulation  period. Pre-whitening the data for the modulation components connected to  one or the other modulation allows us to investigate whether there is any systematic connection between the light-curve variations of the different modulations of a given star. Fig.~\ref{bl2} shows the modulated light curves connected to the different modulation  periods of the double/triple-modulated Blazhko stars. 

Although the utilized light-curve solutions are somewhat uncertain especially for the most complex, and also for the small modulation-amplitude cases, the results are accurate enough to detect any systematic behaviour if it would exist. However, we do not find any trend, connection between the different modulations. There are stars with two modulations of very similar shape but of different  periods (e.g., V079, V054) and also of very different amplitude- and phase-modulation contents like in the case of e.g., V005, V066 and V130. The same is true for the strength of the modulation components, they either have equal/similar strengths (V047), or a large-amplitude, dominant modulation is accompanied by a marginal secondary one (V033, V135).

There is no tendency in the modulation properties with increasing pulsation  periods (from the top to the bottom in Fig.~\ref{bl2}), and with increasing modulation  periods of a given star (from the left to the right in each row of Fig.~\ref{bl2})  either.

Concerning the modulation  periods, they ratios are close to small-integer ratios in some cases (e.g., V005 3:4; V066 3:4; V079 2:5; V130 1:2; V135 1:2), however, the limited length of the data set does not allow  to determine the longer modulation  periods   accurately enough for making any conclusion from these data. We note, however, that a similar tendency was noticed in the study of Blazhko stars in M5 \citep{m5} and in the analysis of the double-periodic modulation of CZ Lac \citep{czl}.

\section{Comparison of Blazhko and stable light-curve RRab stars}

\begin{figure}
\includegraphics[width=8.cm]{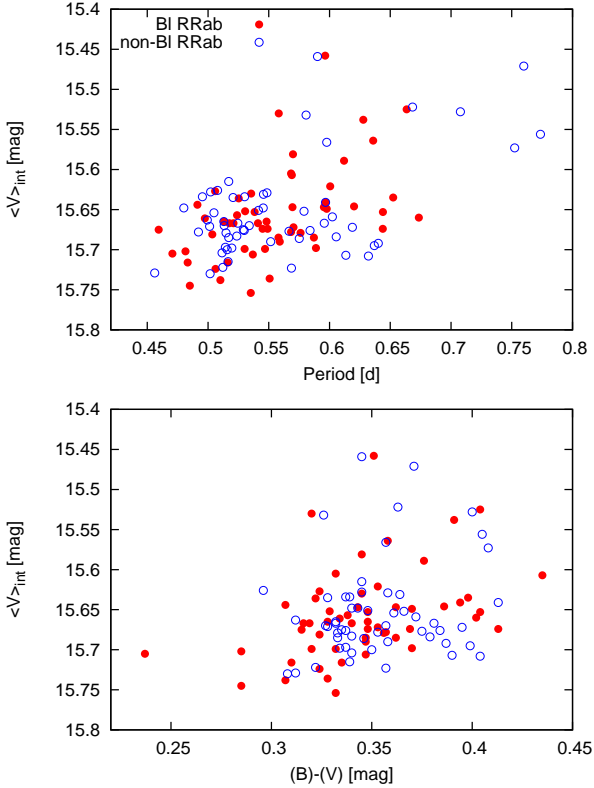}
\caption{Positions of Blazhko and non-Blazhko stars in the $<V>_{\mathrm int}$-period and $<V>_{\mathrm int}$-$(B)-(V)$ colour plots. Blazhko and non-Blazhko RRab stars are denoted by filled- and open-circle symbols, respectively.} 
\label{v-bv}
\end{figure}

$<V>_{\mathrm{int}}$-period and $A$-period plots for Blazhko and non-Blazhko RRL stars in M3 were already shown in \cite{m3mod} and \cite{m3data}. For completeness, here we also document that there is no significant difference between their distributions.
With the exception of the longest-period region, Blazhko stars populate the same parameter ranges as stable light-curve RRab stars in the $<V>_{\mathrm{int}}$-period and in the  $<V>_{\mathrm{int}}$-$(B-V)$ diagrams shown in Fig.~\ref{v-bv}, indicating that the occurrence of the modulation cannot be connected to any special physical property on the horizontal branch. Using photometric formulae to derive colour, absolute brightness and metallicity for the Galactic bulge RRL stars, \cite{prudil} reached to the same conclusion.

The histograms of the period distributions for modulated and non-modulated stars were documented in figure 4. in \cite{m3mod}. On the average, half of the RRab star in the $0.42-0.67$ period range show the Blazhko effect. In contrast, the light curve of each of the seven longest-period stars is stable. The missing modulation in long-period variables is in accordance with the relative weakness of the modulation in the longest-period ($P>0.62$ d) Blazhko stars. This result is also in line with the drastic decrease of the incidence rate of the modulation at longer than 0.625~d pulsation period in the Galactic bulge \citep{prudil} and with similar tendencies detected in the occurrence of the modulation in other systems as  summarized in \citet[][and references, therein]{marek}.

\begin{table} 
\begin{center} 
\caption{Statistical values for  the period, Fourier parameters and [Fe/H]$_{\mathrm{phot}}$ values for the Blazhko and non-Blazhko RRab stars in M3. The standard error in the last digit of the mean is given in parenthesis.  \label{fou.tab}} 
\begin{tabular}{l@{\hspace{1mm}}l@{\hspace{1mm}}c@{\hspace{1mm}}c@{\hspace{1mm}}c@{\hspace{1mm}}l}
\hline
Param.& Mean & S.dev. & Range & Median &Sample \\
\hline
\multicolumn{6}{l}{full sample of stars: 92 non-Bl and 83 Bl}\\
$P$              &0.565(7) & 0.069 & $0.456 - 0.774$& 0.544&nBl \\
                 &0.560(6) & 0.052 & $0.459 - 0.673$& 0.548&Bl   \\
\multicolumn{6}{l}{$V$ data:  55 non-Bl and 54 Bl}\\
$P$              &0.561(9) & 0.070 & $0.456 - 0.774$ & 0.534&nBl\\
                 &0.556(7) & 0.052 & $0.459 - 0.673$ & 0.549&Bl \\
$<V>_{\mathrm{int}}$ &15.65(1) &0.06&$15.46-15.73$& 15.67&nBl\\
                     &15.66(1) &0.06&$15.46-15.74$& 15.66 &Bl  \\
$<V>_{\mathrm{mag}}$ &15.70(1) &0.07&$15.48-15.80$& 15.71&nBl\\
                     &15.69(1) &0.06&$15.50-15.79$& 15.70  &Bl  \\
$A(V)$           & 0.97(4)&0.26&$0.15-1.25$&1.10  &nBl\\
                 & 0.83(3)&0.22&$0.24-1.23$&0.84   &Bl  \\
$a_1$            & 0.34(1)& 0.09&$0.07-0.45$&0.38 &nBl\\
                 &0.31(1)& 0.07&$0.11-0.43$&0.32   &Bl  \\
$R_{21}$         &0.45(1)& 0.05&$0.17-0.53$&0.46 &nBl\\
                 & 0.42(1)& 0.07&$0.24-0.54$&0.44 &Bl  \\
$R_{31}$         &0.33(1) & 0.07 &$0.02-0.37$&0.35 &nBl\\
                 &0.26(1) & 0.08 &$0.05-0.35$&0.28 &Bl  \\
$\phi_{21}$      &2.34(2) & 0.15&$2.18-2.83$&2.27 &nBl\\
                 &2.37(2) & 0.14&$1.96-2.89$&2.35 &Bl \\
$\phi_{31}$      &5.00(5) & 0.36 &$4.66-6.52$&4.86 &nBl\\
                 &4.96(5) & 0.39 &$3.79-6.41$&4.89   &Bl\\
$\mathrm{[Fe/H]}$&$-1.34(2)$ & 0.16 &$-0.44 - -1.62$ & $-1.37$ &nBl\\
                 &$-1.35(2)$ & 0.11 &$-1.10 - -1.62$ & $-1.37$ &nBl -V202\\
                 &$-1.36(6)$ & 0.41 &$+0.52 - -2.67$ & $-1.42$ &Bl\\
\hline
\end{tabular}
\end{center}
\end{table}

\begin{table*} 
\begin{center} 
\caption{Periods and magnitude- and intensity-averaged mean magnitudes and amplitudes of V119 at different epochs. \label{v119.tab}} 
\begin{tabular}{llcccccccl}
\hline
Year&Period& (B) & <B> & A(B) & (V) & <V> &A(V)& (B)-(V) & Ref.\\
\hline
1992-93& 0.51769& 16.002& 15.901&    1.45 &  15.675& 15.616 &   1.15   & 0.33&\cite{cc}\\
1998  &  0.51768& 16.036& 15.947&    1.36 &  15.728& 15.671 &   1.12   & 0.31&\cite{H05,bj}\\
2009  &  0.51762&    -  &   -    &     -   &  15.688& 15.643 &1.51-0.82 &  - &\cite{oc}\\
2012  &  0.51758& 16.051& 15.969 &1.60-1.05& 15.711& 15.667 &1.32-0.76 & 0.34&\cite{m3data}\\
\hline
\end{tabular} 
\end{center}
\end{table*}
\begin{figure}
\includegraphics[width=7.2cm]{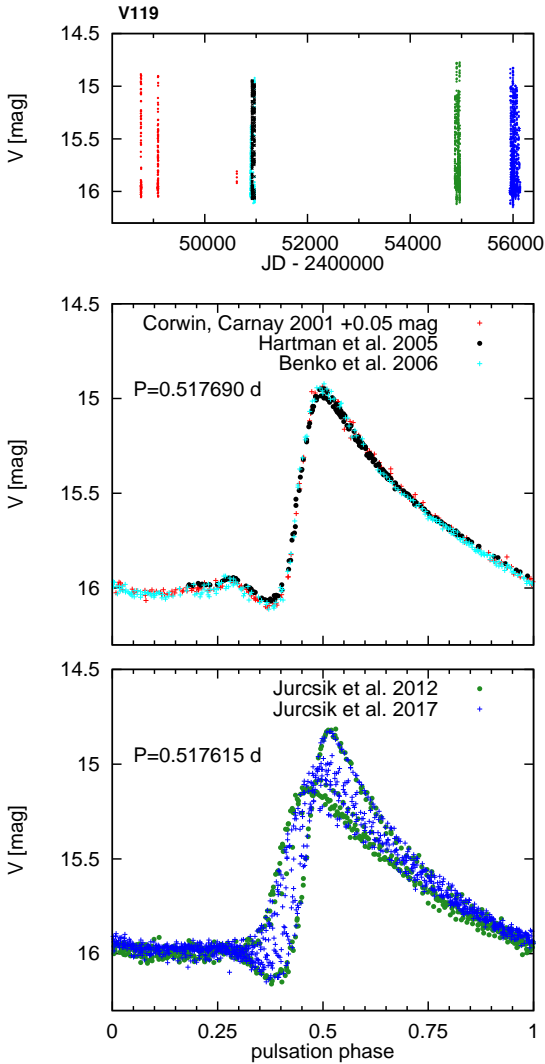}
\caption{CCD light curves of V119. The top panel shows the time distribution of the observations. Phased light curves for the $1992-1998$ and for the $2009-2012$ observations are plotted in the  middle and bottom panels, respectively.
}\label{v119.fig}
\end{figure}

Table~\ref{fou.tab} summarizes the statistics of the different parameters of the light curves for Blazhko and non-Blazhko stars in M3. The parameters correspond to the mean light curve of Blazhko stars, which are somewhat uncertain for variables with incomplete Blazhko-phase coverage.

The difference between the mean periods of Blazhko and non-Blazhko stars is smaller than 0.01 d both for the magnitude calibrated and the total sample of stars, i.e, they are the same within the $1\sigma$ uncertainty limit of the mean. This is in contrast with the statistically significant 0.03 d difference detected between the mean periods of modulated and non-modulated RRL stars in the Galactic bulge \citep{marek}. 

Statistically  significant differences in the mean values derived for the M3 sample are evident only in the amplitudes and amplitude ratios. They are systematically  smaller for Blazhko stars than for  non-Blazhko variables. Contrary to the lowering of the mean epoch-independent phase differences ($\phi_{21}$ and $\phi_{31}$) of the Blazhko sample stars in the Galatic bulge, the mean values of the phase-differences of  the Blazhko and non-Blazhko samples do not differ statistically significantly in M3.

Concerning the parameter ranges, Table~\ref{fou.tab} documents that long-period Blazhko stars are missing, and the phase differences of Blazhko star  cover a wider range  than for stable RRab stars. The small values detected for the amplitudes and amplitude ratios of the non-Blazhko sample corresponds to the very low amplitude, longest-period variable, V202.

The formula given by \cite{jk96} and its different variants are widely used to determine the [Fe/H] values for large samples of RRab stars because  it needs only light-curve information (Fourier parameters) instead of spectroscopic measurements. The photometric metallicity formula are often applied for Blazhko stars, too.  Therefore, it is essential to know how different results are obtained for the [Fe/H] of the Blazhko and non-Blazhko samples of stars in M3, which supposedly have the same metallicity.  Using the \cite{jk96} formula the statistics for the derived  $\mathrm{[Fe/H]_{phot}}$ values are also listed in Table~\ref{fou.tab}.  The mean values of the two samples agree surprisingly well,  although  the the full range of the  $\mathrm{[Fe/H]_{phot}}$  is substantially larger for Blazhko stars than for the non-modulated sample.

\section{The onset of the modulation in V119}

There is a unique variable, V119 in M3. This star switched from regular pulsation to Blazhko-modulated pulsation at the early 2000s \citep{oc}.
Both the stable and the modulated light curves of the star are covered by CCD observations as documented in Fig.~\ref{v119.fig}, and summarised in Table~\ref{v119.tab}. V119 is the only star known where we are witnessing the onset of the modulation in a normal, regular RRab star. This provides a unique possibility to check whether any change in the stellar parameters occurs  in  connection with the onset of the modulation.

Significant changes in the strength of the modulation are detected in other Blazhko stars, too,  e.g., RR Gem \citep{rrg2}, RR Lyr \citep{rrl}, OGLE-BLG-RRLYR-07605 \citep{ogleIV}. However, the modulation does not disappear completely at any time in these stars, only its amplitude diminishes drastically  sometimes. Another similar example is V79 also in M3. A temporal double-mode pulsation was replaced by Blazhko-modulated fundamental-mode pulsation in this star \citep{go10,overtone}.

The periods, mean magnitudes, and amplitudes of V119 determined for the published CCD data are listed in Table~\ref{v119.tab}. These data are derived form the analysis of the published time series, therefore marginal differences from the same results published in the original papers may occur. The \cite{H05} and \cite{bj} observations are overlapping in time, therefore they are merged in order to obtain good phase-coverage light curves. 

The most definite changes detected between the regular and modulated states of the star is the decrease in the pulsation period by 0.00008 d \cite[see also the O-C of V119 shown in][]{oc}. \cite{ogleIV}  showed that the pulsation period of  OGLE-BLG-RRLYR-07605 was also shorter when the amplitude  of the modulation was large than when only small-amplitude modulation appeared. A similar connection betw jurcsik.tgzeen the pulsation period and the amplitude of the modulation was detected in RR Gem, too \citep{rrg2}. However, the historical study of RR Lyrae contradicts these results: its pulsation period and modulation amplitude show parallel changes according to the data listed in Table 3 in \cite{rrl}. 

The changes in the mean-magnitude values are somewhat contradictory. $B$ data are available for three epochs, and these data indicate a continuous dimming of about 0.05 mag,  but no parallel changes are evident in the $V$ band. The $(B)-(V)$ colour might be $0.01-0.02$ mag redder when the star shows the modulation than when its pulsation is stable. The pulsation equation of RRL stars \citep[e.g.,][]{marconi} indicates that a similar order period decrease as observed should be accompanied by luminosity decrease and/or temperature increase significantly below the present-day detection limits. 

We also note that the absolute calibration of the magnitudes of crowded field data might differ by some hundredths of magnitude, therefore the mean-magnitude differences between the different observation of V119 might possibly have photometric origin, i.e.,  they do not necessarily reflect real magnitude changes of the star. If any magnitude change accompanied the onset of the modulation that was probably below the some hundredths of magnitude uncertainty of the accuracy of the photometric calibration of the data sets.

\section{Temperature and radius variations}\label{vi}

\begin{figure}
\includegraphics[width=7.8cm]{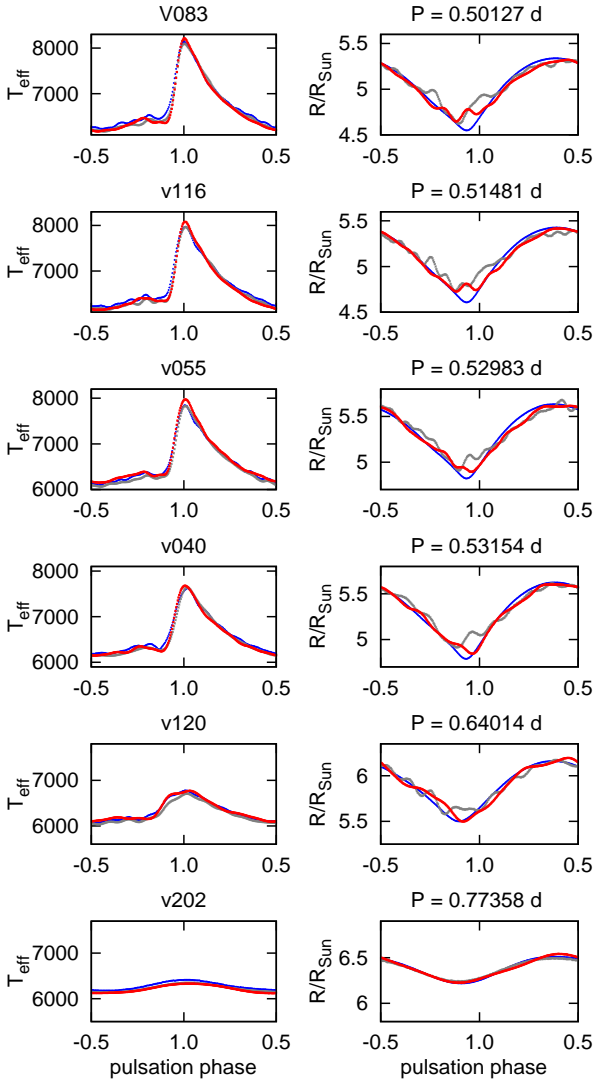}
\caption{Comparison of the radius and temperature changes  of some stable light-curve RRab stars in M3 based on direct Baade-Wesselink method published in \citet{m3data} (blue lines) and applying the photometric method introduced in \citet{vvvbl}  using two-band ($I, K$ - gray lines; $V, I_{\mathrm C}$ -red lines) photometric data and model atmosphere grids \citep{kurucz} and the $R_0$ mean radius values determined by \citet{m3data}.}\label{stable}
\end{figure}

\begin{figure*}
\includegraphics[width=16.8cm]{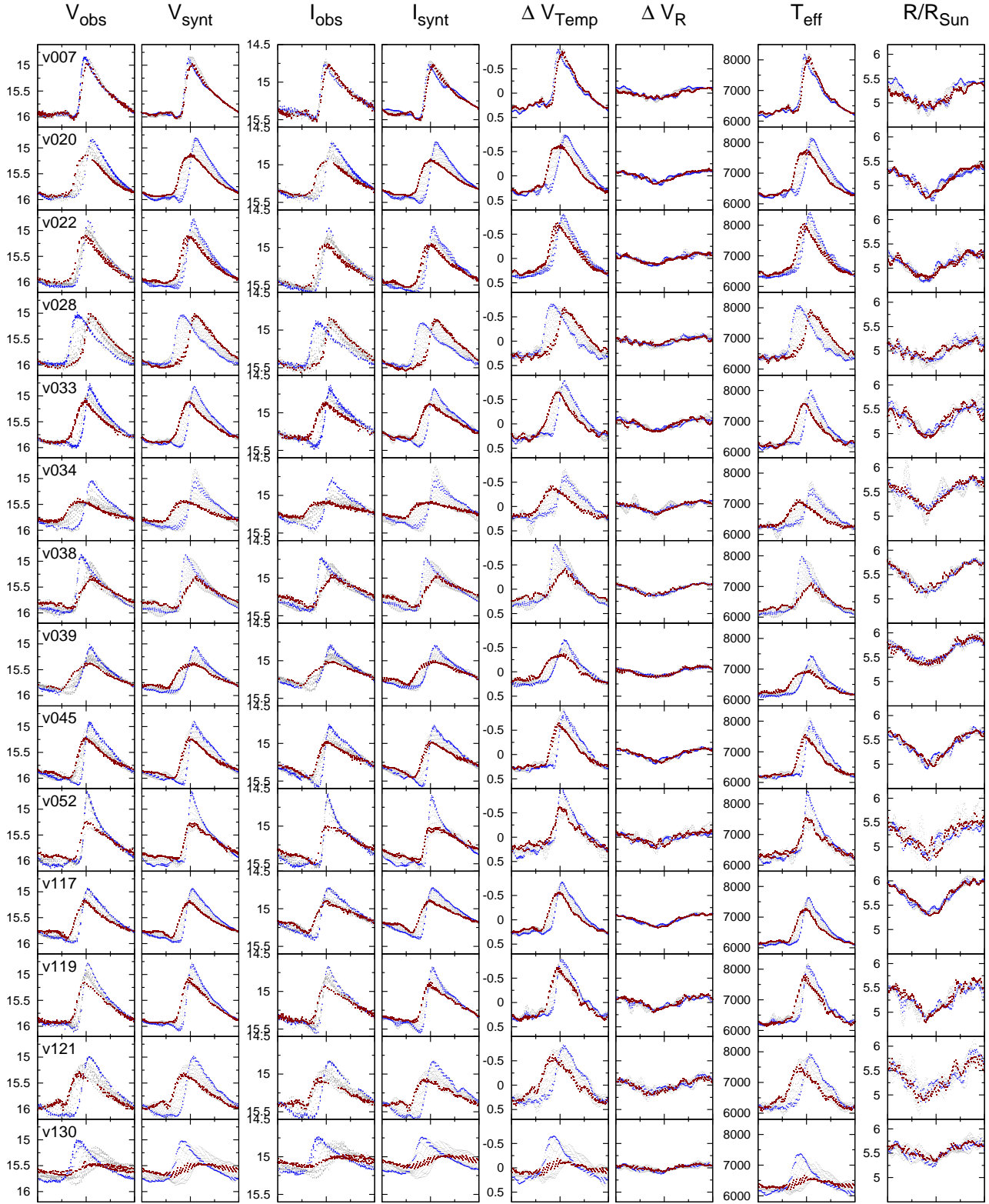}
\caption{The observed and the fitted $V$ and $I_{\mathrm C}$ light curves and the separation of the $V$ light-curve  into temperature- and radius-variation induced changes are shown in the first six columns of the figure for phase-modulated Blazhko stars. The $T_{\mathrm {eff}}$ and $R$ changes are derived from the decomposed V light curves using colour-temperature transformation defined by model atmosphere grids \citep{kurucz} and $R_0$ values determined using the formula given by \citet{marconi}. The observed and the one Blazhko-cycle long simulated light curves are shown by gray dots, and the largest- and the smallest-amplitude phases of the modulation are set out in blue and red colours in each plot. The full description of the method was given in \citet{vvvbl}.
}\label{fmodsep}
\end{figure*}

\begin{figure*}
\includegraphics[width=16.8cm]{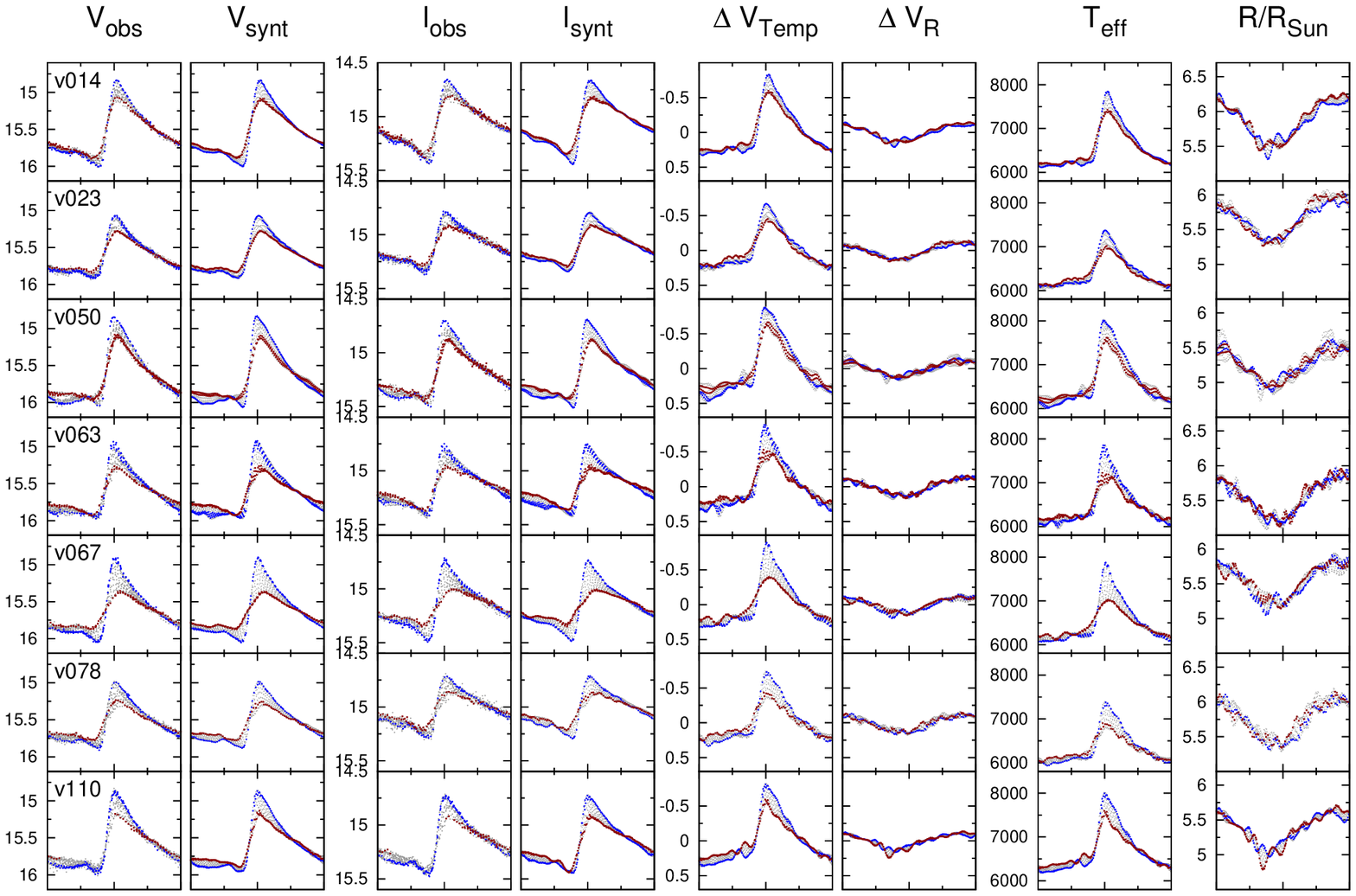}
\caption{The same as Fig.~\ref{amodsep} but for amplitude-modulated Blazhko stars.
}\label{amodsep}
\end{figure*}

The first attempt to perform Baade-Wesselink (B-W) analysis of Blazhko RRL stars has led to a very surprising conclusion \citep{bwbl}. The analysis of the simultaneous photometric and spectroscopic observations of  variables in M3, published in \citet{m3data} showed that the method cannot be applied for Blazhko stars as significant differences between the spectroscopic and the photometric radius  ($R_{\textrm{sp}}$, $R_{\textrm{ph}}$) variations are detected but the B-W method relies basically on the assumption that the $R_{\textrm{sp}}$ and the $R_{\textrm{ph}}$ variations are identical. The phase and amplitude variations of $R_{\textrm{sp}}$ follow the changes of the light curve during the Blazhko cycle but the $R_{\textrm{ph}}$ curve seems to be unaffected or to be only marginally affected by the modulation. A similar conclusion was drawn from the analysis of the combined  OGLE-IV $I$-band  \citep{ogleIV} and the VISTA Variables in the V\'ia L\'actea (VVV, \citealt{vvv}) survey $K$-band observations of  Blazhko stars in the Galactic bulge \citep{vvvbl}. The lack of any phase change of the photosheric radius variations of the pulsation also means that there is no change in the period of the pulsation during the Blazhko cycle, and as a consequence of the period-density relation, the mean radius of the star does not change either.


The method to decompose the $K$-band light variation of RRab stars into two parts, originating  from the temperature and the radius changes developed in \cite{vvvbl} utilizes multi-colour data, colour-temperature relation determined from the \cite{kurucz} synthetic data of appropriate composition, and an estimate on the $R_0$ mean radius value. The $K$-band observations of M3 variables published by \cite{m3k} combined with our photometric data make is possible to check whether a similar method could also be applied for the $V$ and $I_{\mathrm C}$ data. The temperature and radius changes derived from the direct  B-W method, which utilizes both spectroscopic and photometric data, and from the photometric method  using the $K$-band light and $I-K$ colour and the $V$-band light and the $V-I_{\mathrm C}$ colour data are shown for six stable light-curve RRab stars in Fig.~\ref{stable}. The $E(V-I_{\mathrm C}$) reddening is taken to be 0.018 mag \citep[$E(B-V)=0.014$ mag,][]{h96}, uniformly.  The agreement between the three solutions is satisfactory. The radius and temperature changes are reproduced reasonable well when using the $V$ and $I_{\mathrm C}$ colours.

Applying this method on the photometric data  published in \cite{m3data}, the radius and temperature changes are derived for the different phases of the modulation for Blazhko stars. Only variables showing strong and not too complex modulation, and which have good Blazhko-phase coverage are analysed here. The method needs the knowledge of the $R_0$ mean radius value of the stars, which is calculated according to the log$R/R_{\odot}$($\log p$, $\log Z$) formula derived by \cite{marconi}. 

 The light curves and the derived variations of the $T_{\mathrm{eff}}$ and $R$ curves for Blazhko stars showing significant phase modulations are shown in Fig.~\ref{fmodsep} and for amplitude-modulated variables in Fig.~\ref{amodsep}, respectively. The wiggles seen especially in the radius curves originate from the incompleteness of the synthetic data for long Blazhko cycles (e.g., V034, V052) and from the magnification of the small wiggles on the synthetic colour curves  when transforming the logarithmic magnitude scale to linear.

\begin{figure}
\includegraphics[width=7.6cm]{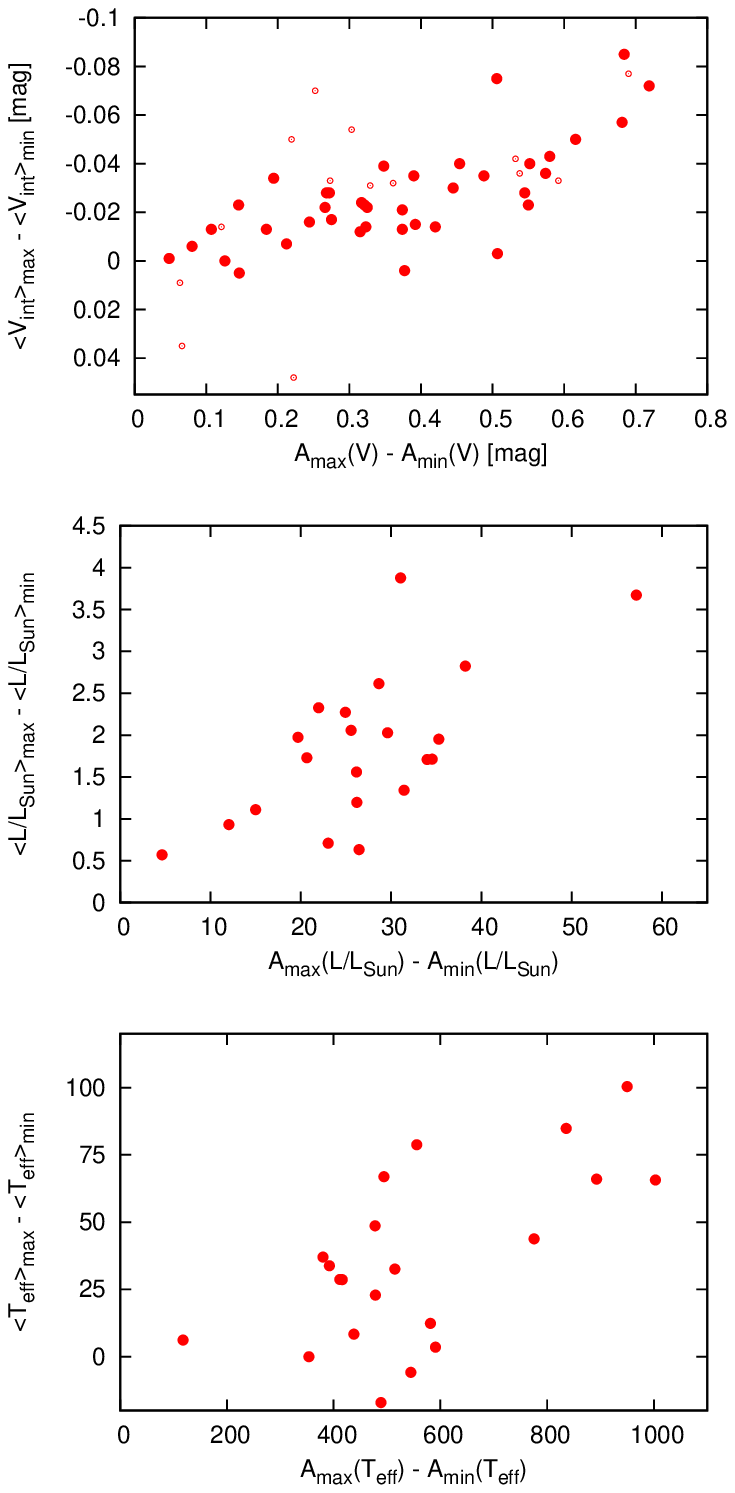}
\caption{Changes in the intensity-averaged mean-$V$ magnitudes versus the $V$-amplitude change between the largest- and smallest-amplitude phases of the modulation are plotted for all the magnitude-calibrated variables in the top panel. Stars with incomplete Blazhko-phase coverage (long Blazhko cycle) are shown by small symbols. The middle and bottom panes show the changes of the mean luminosity and mean temperature values versus their amplitude change, respectively. These two  plots are shown only for the variables analysed in Section~\ref{vi}.
}\label{meanvar}
\end{figure}

The results shown in Figs.~\ref{fmodsep} and ~\ref{amodsep}  are in accordance with our previous findings. They show that no or marginal photometric radius change is behind the observed light-curve modulation, i.e., this reflects exclusively/basically temperature changes.

\section{Changes of the mean parameters}

 In the course of the analysis of the Konkoly Blazhko Survey data \citep{stat} an inverse photometric (IP) B-W method was developed \citep{ip} aiming to detect the changes of the mean physical parameters of the variables in different Blazhko phases. The application of the method revealed  that the mean luminosity of the stars was larger by $1-2$ percent in the large-amplitude phase of the modulation  than in the small-amplitude phase. Each of the stars analysed by the IP method followed this pattern \citep[][and references therein]{2rr}.  The failure of the direct B-W method on Blazhko RRL stars \citep{bwbl} warns, however, that the results obtained from the IP method might be false.

The analysis performed in Section~\ref{vi} yields an other estimate for the  changes of the mean parameters in different Blazhko phases. The differences between the mean parameter values observed at Blazhko maximum and minimum phases are determined and it is checked whether they show any connection with the direction and strength of the parallel amplitude change. This is done by plotting the detected differences of the mean values against the amplitude change between the two extrema of the modulation.

Fig.~\ref{meanvar} shows the results for the intensity-averaged mean $V$ magnitudes of all the magnitude-calibrated Blazhko stars (top panel), and for the mean luminosity (middle panel) and mean temperature (bottom panel) changes between the largest- and smallest-amplitude phases for the 21 variables analysed in Section~\ref{vi}. The luminosity curves are calculated using the derived radius- and temperature-change curves. They are not shown in Figs.~\ref{fmodsep} and ~\ref{amodsep}, as they are very similar to the $V$ light curves. We note here again, that the mean radius values are fixed for each star, however the lack of any period(phase) change of the radius variations validate this treatment. 

Fig.~\ref{meanvar} proves that the changes of the intensity-averaged mean $V$ magnitude, as well as the mean luminosity and the mean temperature correlates with the amplitude change. Blazhko  stars tend to be  brighter and hotter at the large-amplitude phase of the modulation than in the small-amplitude phase. Moreover the larger the amplitude of the modulation the larger the changes of the mean parameters.

\section{Summary and conclusions}

The analysis of the time-series data of the total sample of  Blazhko stars in the M3 globular cluster, i.e., in a homogeneous group of variables, makes it possible to detect any connection between the pulsation and modulation properties, if any direct relationship would exist. Moreover, the comparison of the results with the data of the single-periodic RRL sample of the cluster yields a unique opportunity to detect any systematic difference between regular and Blazhko RRL stars. However, most of these studies led to  basically negative conclusions.

The results can be summarised as follows:
\begin{itemize}

\item{Comparing the mean log$(P_{\mathrm{mod}})$ values of Blazhko stars in M3 and  in the sample/subsamples collected by \cite{marek} we have found that the mean Blazhko periods are larger in more metal-poor systems than in less metal-poor ones.} 

\item{The only regularity revealed by the analysis of M3 Blazhko stars is that the modulation of long-period Blazhko stars ($P>0.62$ d) is weak, and this is also valid when the relative strength of the modulation ($\Delta A/<A>$) is considered.
This is in line with the lack of the modulation in the seven longest-period  ($P>0.66$ d) stars in the cluster. This is the only systematic difference between the distributions and the properties of Blazhko and regular RRab  stars in the cluster.}

\item{Another systematic feature supposed to be found is the lack of intense phase modulation in  OoII-type stars. Nevertheless, because of the limited size of the  sample this conclusion might not be true in general.}

\item{The light curves of Blazhko stars showing small-amplitude modulation are close to the normal ones, and strongly deformed-shape light curves are detected only if the modulation is strong. }

\item{The mean amplitudes of Blazhko stars are significantly smaller than the  normal ones for variables exhibiting strong modulation independently from that the phase or the amplitude modulations or both are intense. The maximum amplitudes are either smaller or larger or similar to normal amplitudes for strongly modulated light curves.}

\item{Besides the above mentioned tendencies, the data did not show up any well-determined, direct connection between the physical properties of the stars and their Blazhko characteristics (modulation period, amplitude, phase/amplitude modulation strengths). No regularity was detected in the modulation properties of multiple modulated stars either. With the exception of  the longest-period (coolest) variables, the modulation occurs all along the instability strip, and a non-modulated counterpart was found for most of the Blazhko stars in the M3 sample.}

\item{Modulation appeared in a previously stable RRab star (V119)  in the early 2000s.  Parallel with the onset of the modulation,  a 0.00008-d pulsation-period decrease of the star was observed. Rapid/sudden period changes of the same order are not rare in RRL stars, and Blazhko stars may exhibit even larger period-change events of unknown origin  \citep[for possible explanations see e.g.,][]{sr79,st80,k94,cox}. The parallel magnitude/colour changes of V119 do not lead to any conclusive result. They originate from photometric inaccuracies probably. }

\item{Although the [Fe/H]$_{\mathrm{phot}}$ values derived for Blazhko stars spread over a much wider range than for non-Blazhko stars, the mean [Fe/H]$_{\mathrm{phot}}$ values of the two samples are identical indicating that the photometric method yields a reliable estimate for the mean metallicity of the system for large enough samples.}

\item{The separation of the temperature- and radius-change induced variations of the $V$ light curve confirms our previous finding that the radius variation of the  photospheric regions does not show any detectable phase or amplitude change parallel with the observed light-curve variations of Blazhko stars.}

\item{Both the intensity-averaged mean $V$ brightness, the mean temperature and the mean luminosity tend to be brighter/larger in the large-amplitude phase of the modulation than in the small-amplitude phase. The larger the amplitude of the modulation, the larger changes in the mean parameters are detected.  }

\end{itemize}


Based on these findings the following conclusions  can be drawn. $a)$ 
{\ With the exception of the coolest variables in the vicinity of the red edge of the instability strip, the occurrence and the properties of the modulation do not depend on the physical parameters of the variables. This is in high contrast with the properties of the radial mode pulsation, which is determined by the physical parameters of the stars completely.} It is to be hoped only that the modulation properties averaged over large samples of Blazhko stars of different populations may show some systematic differences as indicated by the proposed metallicity dependence of the  mean  log$(P_{\mathrm{mod}})$ value. $b)$ The similarity of modulated and non-modulated stars and the appearance of the modulation in V119, and other examples of drastic changes in the modulation amplitudes e.g., the drop of the modulation amplitude of OGLE-BLG-RRLYR-07605 \citep{ogleIV}, taking into account also the temporal disappearance/lowering of the modulation in RR Lyr \citep{rrl,poretti} and in RR Gem \citep{rrg2} and the strange case of V79 in M3 \citep{overtone} indicate that the Blazhko modulation is not a stable feature of the pulsation of RRL stars, it may appear or disappear in any RRL star at any time, probably.  $c)$ Whatever is the physical mechanism of the Blazhko effect, the weakening of the relative strength of the modulation and the lack of any modulation in the longest-period variables indicate that maybe the enhanced convection close to the red edge of the instability strip stops modulation prior to halting the pulsation itself. Alternatively, the evolutionary state or the cool temperature of variables close to the red edge of the instability strip does not support the occurrence of the modulation. The $0.62-0.65$ d period-length limit for the modulation seems to be valid in different groups of RRL stars as noted by \cite{prudil}.   $d)$ As a consequence of the stability of the photospheric radius changes, the pulsation period has to remain stable during the Blazhko cycle.
$e)$ The energy of radiation emission averaged over the pulsation period is larger in the large-amplitude Blazhko phase than in Blazhko minimum. Most probably,  the same is true for the changes of the pulsation averaged kinetic energy, as the large amplitude changes of the  $R_{\textrm{sp}}$ data indicate.  As a consequence, we conclude that some additional source of energy storage/release should have to operate during the Blazhko cycle.

Based on the spectroscopic and photometric observations obtained in the last decade several new features of the Blazhko modulation has been discerned. However, all these efforts has not led to a breakthrough in solving the Blazhko mystery yet. Therefore, it is the task primarily of theorists  now to find a model fitting to all of recent knowledge on the Blazhko phenomenon.

\section{Acknowledments}
The support of the National Research, Development and Innovation Office  NN-129075 and K-129249 grants is acknowledged. The author is grateful to J\'anos Nuspl for fruitful discussions and for his comments on the manuscript. Remarks and suggestions of Gergely Hajdu helped to inprove the paper.
 
{}

\appendix
\section{List of Blazhko stars and their periods}\label{lck}

\begin{table*}
 \begin{minipage}{\textwidth}
\begin{center} 
\caption{Summary of the periods of the Blazhko RRab stars in M3.\label{rrbl.dat}} 
\begin{tabular}{llllll}
\hline
Star/Oo~type& P$_{\mathrm{puls}}$ &  P$_{\mathrm{Bl}}$ & P$_{\mathrm{puls}}$ &  P$_{\mathrm{Bl}}$& Comment \\ 
                         &\multicolumn{2}{c}{{\hskip -35pt}2012}& \multicolumn{2}{c}{{\hskip -25pt}2009-2012}&\\
\hline
V003 OoII  & 0.55820 &\hskip 4pt 37.5       &         &              &  \\
V004b       & 0.59305 &\hskip 4pt 96.1       &         &              &  \\  
V005 OoI   & 0.50579 &\hskip 4pt 76.7/103  &         &              & \\
V007 OoI   & 0.49742 &\hskip 4pt 51.9       &         &              &  \\ 
V008        & 0.63671 &      109           &         &              &  \\  
V010 OoI   & 0.56955 &      191           & 0.56955 & 189.2        &  \\ 
V014 OoII  & 0.63590 &\hskip 4pt 60.6       &         &              &  \\ 
V017 OoI   & 0.57613 &\hskip 4pt 83.3/132 & 0.57615 &\hskip 4pt  82/160  &  \\ 
V020 OoI   & 0.49126 &\hskip 4pt 44.2       &         &              &  \\ 
V022 OoI   & 0.48143 &      140           &         &              &  \\ 
V023 OoI   & 0.59538 &      125           &         &              &  \\ 
V024 OoII  & 0.66343 &\hskip 4pt 49.8       &         &              &  \\ 
V028 OoI   & 0.47066 &\hskip 4pt 65.9       &         &              &  \\ 
V033 OoI   & 0.52524 &\hskip 4pt 40.5/34.8:       &         &          &  \\ 
V034 OoI   & 0.55901 &      300:            & 0.55909 & 305.4        & extra frequency at 2.605500 d$^{-1}$  \\ 
V035 OoI   & 0.53058 &\hskip 4pt 56.2/150:  & 0.53059 &\hskip 4pt 57.1/147     &  \\ 
V038 OoI   & 0.55801 &\hskip 4pt 51.8       &         &              &  \\   
V039 OoI   & 0.58710 &      200             & 0.58707 & 190.3        &  \\ 
V041 OoI   & 0.48504 &      190           & 0.48508 & 197.8        &  \\ 
V043        & 0.54053 &      103           &         &              &  \\ 
V045 OoI   & 0.53691 &\hskip 4pt 26.6       &         &              &  \\ 
V047 OoI   & 0.54110 &\hskip 4pt 69.0/53.3/44.6        &  &        &  \\ 
V048 OoI?  & 0.62783 &      143           & 0.62783 & 142.7        & see comment in \cite{bwbl} \\ 
V049 OoI   & 0.54821 &      400:            & 0.548206 & 400.7       &  \\  
V050 OoI   & 0.51309 &      178:            & 0.51310 &  177.5       &  \\ 
V052 OoI   & 0.51624 &      195:            & 0.516234 & 186.3       & 1998-2012$^{*}$\\ 
V054 OoI   & 0.50613 &\hskip 4pt 57.8/80.0    &         &            &  \\ 
V059 OoI   & 0.58882 &\hskip 4pt 62.1       &         &              &  \\  
V061 OoI   & 0.52091 &      134:            & 0.52089 & 175.9        & $f_1=2.5806$ d$^{-1}$   \\  
V062 OoI   & 0.65240 &      270:            & 0.65240 & 302.4        &  \\ 
V063 OoI   & 0.57040 &\hskip 4pt 58.7       &         &              & \\ 
V066 OoI   & 0.62015 &\hskip 4pt 65.9/49.3  & 0.62015 &\hskip 4pt   66.3  49.3 &  \\ 
V067 OoI   & 0.56832 &      103           &         &              & \\ 
V071 OoI   & 0.54905 &\hskip 4pt 46.7       &         &              &  \\ 
V073 OoI   & 0.67349 &\hskip 4pt 54.3       &         &              &   \\ 
V077 OoI   & 0.45935 &      116           &         &              &  \\ 
V078 OoII? & 0.61192 &\hskip 4pt 39.5       &         &              & \\ 
V079 OoI   & 0.48329 &      160/62.9/18.3:        &         &              & \\ 
V080 OoI   & 0.53845 &      300:            & 0.53846 &  305.7       & \\ 
V091 OoI   & 0.53013 &      400:            & 0.53011 &  395.0       & \\ 
V092 OoI   & 0.50355 &      300:            & 0.50355 &  312.8       & \\ 
V101 OoI   & 0.64389 &      103           &         &              &   \\ 
V104 OoII  & 0.56993 &      110           & 0.56993 &  107.2       & \\ 
V106 OoI   & 0.54689 &\hskip 4pt 63.6       &         &              &   \\  
V110 OoI   & 0.53547 &\hskip 4pt 44.9       &         &              &   \\ 
V111 OoI   & 0.51019 &\hskip 4pt 78.2       &         &              &   \\ 
V114 OoI   & 0.59773 &\hskip 4pt 54.5       &         &              &   \\  
V117 OoI   & 0.60054 &\hskip 4pt 47.2       &         &              &   \\  
V119 OoI   & 0.51758 &\hskip 4pt 56.9       &         &              &   \\  
V121 OoI   & 0.53520 &\hskip 4pt 65.1       &         &              &   \\  
V130 OoI   & 0.56899 &\hskip 4pt 85.1/195:  &         &              &  \\ 
V133 OoI   & 0.55073 &      152:            &         &              &  \\  
V134        & 0.61805 &\hskip 4pt\,\,8.4:   &         &              & modulation uncertain \\
V135 OoI   & 0.56840 &\hskip 4pt 63.5/133:       &         &              & \\
V143 OoII  & 0.59641 &      250:            &         &              &  \\ 
V144 OoI   & 0.59670 &   --               & 0.596791 & $>$9000        & 1992-2012$^{**}$\\ 
V149        & 0.54816 &      160:          &         &              & \\   
V150 OoI   & 0.52392 &\hskip 4pt 24.7     &         &              &  \\ 
V151        & 0.51682 &      250:          &         &              &  \\              
V157        & 0.54285 &\hskip 4pt 51.8     &         &              &  \\            
V159        & 0.53382 &      100           &         &              & \\  
V160        & 0.65730 &\hskip 4pt 30.8     &         &              &  \\       
\end{tabular}
\end{center}
\end{minipage}
\end{table*}

\begin{table*}
\begin{minipage}{\textwidth}
\begin{center}
\contcaption{}
\begin{tabular}{llllll}
\hline
Star/ Oo~type& P$_{\mathrm{puls}}$ &  P$_{\mathrm{Bl}}$ & P$_{\mathrm{puls}}$ &  P$_{\mathrm{Bl}}$& Comment \\ 
                &\multicolumn{2}{c}{{\hskip -25pt}2012}& \multicolumn{2}{c}{{\hskip -0pt}2009-2012}&\\
\hline
V161        & 0.52656 &\hskip 4pt 43.4     &         &              &  \\ 
V165        & 0.48363 &\hskip 4pt 20.3     &         &              &  \\          
V167 OoI   & 0.64395 &\hskip 4pt 65.7     &         &              &  \\ 
V174        & 0.59123 &\hskip 4pt 80.7     &         &              &  \\         
V176        & 0.53959 &      145:          &         &              &  \\        
V184        & 0.53128 &\hskip 4pt 45.6     &         &              &  \\       
V186        & 0.66327 &\hskip 4pt 57.7     &         &              &  modulation uncertain\\             
V191        & 0.51918 &\hskip 4pt 41.1/28.9  &         &              &  \\     
V192        & 0.49727 &\hskip 4pt 82.9     &         &              & \\              
V195        & 0.64397 &\hskip 4pt 56.7     &         &              & \\          
V201        & 0.54052 &      210:          &         &              &  \\ 
V211        & 0.55820 &\hskip 4pt 69.3     &         &              & \\ 
V212        & 0.54222 &      230:          &         &              & modulation uncertain, extra frequency at 4.727706  \\
V214        & 0.53953 &\hskip 4pt 93.9     &         &              &  \\ 
V218 OoI   & 0.54487 &\hskip 4pt 79.6     &         &              &  \\ 
V219        & 0.61363 &\hskip 4pt 56.4     &         &              &  \\ 
V220        & 0.60016 &\hskip 4pt 43.6     &         &              &  \\ 
V239        & 0.50399 &\hskip 4pt 77.8     &         &              &  \\ 
V243        & 0.63462 &\hskip 4pt 45.0     &         &              &  \\              
V255        & 0.57265 &      155:          &         &              &  \\              
V270b       & 0.62585 &\hskip 4pt 75.0     &         &              &  \\
\hline 
\multicolumn{6}{l}{$^*$ Additional data from \cite{bj} and \cite{oc}.}\\
\multicolumn{6}{l}{$^{**}$ Additional data from \cite{cc}, \cite{bj} and \cite{oc}.}
\end{tabular}
\end{center}
\end{minipage}
\end{table*}

\end{document}